\documentclass[sigconf, nonacm]{acmart}

\usepackage{amsmath}
\usepackage{amsfonts}
\usepackage{algorithmic}
\usepackage{graphicx}
\usepackage{pgfplots, pgfplotstable}
\usepackage{textcomp}
\usepackage{booktabs}
\usepackage{url}
\usepackage{tabularx}
\usepackage{threeparttable} 
\usepackage{cleveref}
\usepackage{enumitem}
\usepackage[english]{babel}
\def\BibTeX{{\rm B\kern-.05em{\sc i\kern-.025em b}\kern-.08em
    T\kern-.1667em\lower.7ex\hbox{E}\kern-.125emX}}

\begin{document}
\title[]{Evaluating the Impact of Data Cleaning on the Quality of Generated Pull Request Descriptions}

\author{Kutay Tire}
\affiliation{%
  \institution{Bilkent University}
  \department{Department of Computer Engineering}
  \city{Ankara}
  \country{Turkey}
}
\email{kutay.tire@ug.bilkent.edu.tr}

\author{Berk Çakar}
\affiliation{%
  \institution{Purdue University}
  \department{Department of Electrical and Computer Engineering}
  \city{West Lafayette}
  \country{USA}
}
\email{bcakar@purdue.edu}

\author{Eray Tüzün}
\affiliation{%
  \institution{Bilkent University}
  \department{Department of Computer Engineering}
  \city{Ankara}
  \country{Turkey}
}
\email{eraytuzun@cs.bilkent.edu.tr}

\begin{abstract}
\textbf{Background:} Pull Requests (PRs) are central to collaborative coding, summarizing code changes for reviewers. However, many PR descriptions are incomplete, uninformative, or have out-of-context content, compromising developer workflows and hindering AI-based generation models trained on commit messages and original descriptions as ``ground truth.''\\
\textbf{Aim:} This study examines the prevalence of ``noisy'' PRs and evaluates their impact on state-of-the-art description generation models.\\
\textbf{Method:} We propose four cleaning heuristics to filter noise from an initial dataset of 169K+ PRs drawn from 513 GitHub repositories. We train four models—BART, T5, PRSummarizer, and iTAPE—on both raw and cleaned datasets. Performance is measured via ROUGE-1, ROUGE-2, and ROUGE-L metrics, alongside a manual evaluation to assess description quality improvements from a human perspective.\\
\textbf{Results:} Cleaning the dataset yields significant gains: average F1 improvements of 8.6\% (ROUGE-1), 8.7\% (ROUGE-2), and 8.5\% (ROUGE-L). Manual assessment confirms higher readability and relevance in descriptions generated by the best-performing model, BART when trained on cleaned data.\\
\textbf{Conclusion:} Dataset refinement markedly enhances PR description generation, offering a foundation for more accurate AI-driven tools and guidelines to assist developers in crafting high-quality PR descriptions.
\end{abstract}


\maketitle

\section{Introduction} \label{intro}
Pull requests, commonly referred to as PRs, serve as a means for developers to view the changes they have pushed to a repository on code hosting platforms like Bitbucket or GitHub \cite{Gousios2014}. Typically, a PR consists of one or more interrelated commits. To create a PR on GitHub, a developer must provide a title and has the option to include a PR description to provide further details about the changes made. The PR description can be particularly useful for reviewers as it allows them to quickly understand the context of the changes without needing to dive deeply into the details. 
Yet, developers often neglect to write a PR description that helps reviewers understand the changes made or mentions a solved issue, despite the fact that including a description can reduce the chance of PRs being rejected \cite{fan2018early}.

Despite the significance of including a PR description, its mere presence may not be sufficient. The quality of the provided description also holds significant weight, influencing the likelihood of a PR being accepted \cite{gousios2015work, fan2018early}. In light of this idea, Yu et al. \cite{Yu2015} demonstrate that PRs with high-quality descriptions may experience reduced evaluation latency, meaning that the time reviewers take to assess and approve such PRs is often shorter. This underscores the practical impact of well-crafted PR descriptions on the efficiency of the review process.

Expanding on this idea, there has been a notable increase in research on the automatic generation of PR titles and descriptions, focusing on developing models that can summarize a given input sequence consisting of artifacts such as commit messages to obtain a description or title. While some new models have been developed specifically for description generation, such as PRSummarizer \cite{liu2019automatic} or PRHAN \cite{fang2022prhan}, existing pre-trained models like BART \cite{lewis2019bart} and T5 \cite{raffel2019exploring} have also been employed for this purpose \cite{zhang2022automatic}. However, these models were trained with the PRs from open-source projects while counting on the descriptions provided by the developers as the \textit{reference description} or namely the \textit{ground truth description} \cite{liu2019automatic, fang2022prhan, zhang2022automatic}. 

Since these descriptions are initially authored by developers, they might be subjective, potentially including personal feelings or plans, or they may contain uninformative content that does not effectively summarize the PR. Similarly, the commit messages associated with a PR can also be subjective or lack informative content about the proposed changes. Consequently, the quality of PRs with such ground truth descriptions or commit messages becomes a concern, and these PRs are assumed to be ``noisy'' as they can negatively impact the descriptions produced by the generation models. Supporting this notion, Tuzun et al. \cite{tuzun2022ground} suggest that potential cognitive biases of individuals could substantially impact the performance of automated models in software engineering by influencing the ground truth. Therefore, the same issue may arise for description generation models, where the generated descriptions could exhibit subjectivity and include irrelevant or uninformative content rather than focusing on specific code modifications. 

To solve this problem, we propose four cleaning heuristics to eliminate noise in PR description datasets. The first heuristic focuses on eliminating trivial commit messages from the dataset, while the second heuristic targets trivial PR descriptions. Both of these heuristics are designed to remove content that does not provide any informative value about the PR. Among the remaining, one heuristic eliminates PR samples with irrelevant descriptions, such as those containing subjective or out-of-context information, and another eliminates PR samples if there is a significant mismatch between the lengths of their descriptions and commit messages, as this would affect the length of the generated descriptions and performance of the models. After applying our heuristics to the initially collected dataset, we obtain a cleaned benchmark dataset for the PR description generation task.

To evaluate the effectiveness of the heuristics, we use four state-of-the-art text summarization models, including both domain-specific and general-purpose summarization models. We choose all four of these models to be abstractive, capable of generating sentences that are different from the original sentences in the input sequence, to better understand the effect of cleaning. We use BART \cite{lewis2019bart} and Text-To-Text Transfer Transformer (T5) \cite{raffel2019exploring} as general-purpose summarization models, both of which are large pre-trained models. For domain-specific summarization models, we employ PRSummarizer \cite{liu2019automatic} and iTAPE \cite{chen2020stay}. 


In this study, we defined the input sequence as the concatenation of commit messages. The target sequence, on the other hand, is the generated PR description. To construct our dataset, we applied the necessary preprocessing to the initially collected PRs and obtained 625K+ PRs from 513 GitHub repositories. After splitting our dataset into training, validation, and test sets, we applied our cleaning heuristics to each set to eliminate noise. As a result, we obtained a cleaned dataset with a total of 169K+ high-quality PR samples.

To understand how our cleaning heuristics affect the quality of the generated descriptions and the performance of the text summarization models, we have formulated two research questions:

\textbf{RQ1:} \textit{How does the application of our cleaning heuristics to PR description datasets affect the performance of the summarization models?}

\textbf{RQ2:} \textit{To what extent do these heuristics effectively identify noisy PR samples and contribute to the generation of better-quality descriptions from a human perspective?}

For RQ1, to investigate the effects of our cleaning heuristics, we train the summarization models separately on the initial uncleaned dataset and the cleaned dataset. Subsequently, we compare the precision, recall, and F1 scores of each model on the same test set for both cases using the ROUGE metrics \cite{lin2004rouge}, which are standard metrics for the automatic evaluation of summarization approaches.

For RQ2, we conduct a manual evaluation consisting of two stages. In the first stage, two authors of this study independently compare descriptions generated from BART \cite{lewis2019bart}, the best-performing model, before and after the model is trained on the cleaned dataset. In the second stage, for each heuristic, the PRs marked as noisy are analyzed and labeled as either \textit{true positive} or \textit{false positive} to assess whether these heuristics successfully identify the noisy PR samples.

In summary, our contributions are as follows:

\begin{itemize}
    \item We propose four cleaning heuristics to refine samples for training and evaluating the text summarization models (BART \cite{lewis2019bart}, T5 \cite{raffel2019exploring}, PRSummarizer \cite{liu2018neural}, iTAPE \cite{chen2020stay}).  The results show that the models' F1 scores increase by 8.6\%, 8.7\%, 8.5\% for ROUGE-1, ROUGE-2, and ROUGE-L metrics, respectively, on average after applying our heuristics.
    \item We build a noise-free dataset with over 169K+ PRs from GitHub for the task of automatic PR description generation.
    \item We emphasize the analysis of generated descriptions from a human perspective to ensure their relevance and quality. To achieve this, we perform a comprehensive manual assessment that evaluates the effectiveness of each heuristic individually and examines their combined impact on the overall quality of the descriptions.
\end{itemize}

The sections of this paper are organized as follows: \Cref{background-sec} introduces the motivating examples. \Cref{meth-sec} focuses on the methodology, detailing the specifics of our cleaning heuristics and text summarization models. \Cref{eval-sec} details the manual evaluation setup and metrics used for automatic evaluation. \Cref{setup-sec} describes the experimental settings and \Cref{results-sec} presents the results of our experiments. \Cref{validity-sec} discusses the threats to validity, outlining the potential limitations and biases in our approach, while \Cref{disc-sec} explores the potential implications of our work, highlighting its relevance and possible impact on future research and practice. After a review of related work in \Cref{rel-work-sec}, we conclude the paper with \Cref{conc-sec}.

\newcolumntype{C}{>{\arraybackslash}X} 
\begin{table*}[!ht]
\footnotesize
\centering
\caption{List of Poorly Written PR Samples}
\setlength\extrarowheight{2pt} 
\resizebox{0.97\linewidth}{!}{%
\begin{tabularx}{\textwidth}{|C|C|C|C|}
\hline
\textbf{Example 1} & \textbf{Example 2} & \textbf{Example 3} & \textbf{Example 4}\\
\hline
\textbf{Source:} \url{https://github.com/AltBeacon/android-beacon-library/pull/578} &
\textbf{Source:} \url{https://github.com/pytorch/pytorch/pull/21330} &
\textbf{Source:} \url{https://github.com/javaparser/javaparser/pull/470} &  
\textbf{Source:} \url{https://github.com/t-oster/VisiCut/pull/387} \\
\textbf{Original PR Description:} \textit{As described in \#577, in some cases a ConcurrentModificationException will cause an app using the library to crash if it modifies the monitored regions at the same time the app begins a passive scan. This change fixes that.} & 
\textbf{Original PR Description:} \textit{fix issue \#21271} &
\textbf{Original PR Description:} \textit{Awful formatting problem I know.. :/} &
\textbf{Original PR Description:} \textit{This is the first draft for \#384 . It does introduce some dependencies to the VisiCut model in the Code and is also not very well designed, but I think it works. Happy testing.} \\
\textbf{Commit Messages:} 
\begin{itemize}[leftmargin=*, align=left]
  \item \textit{Fix ConcurrentModificationException starting passive scan per \#577}
  \item \textit{Merge branch 'master' into fix-cme-on-start-passive-scan}
  \item \textit{Update changelog for fixing CME on Android 8 passive scan start}
  \item \textit{Merge branch 'master' into fix-cme-on-start-passive-scan}
\end{itemize} &
\textbf{Commit Messages:} 
\begin{itemize}[leftmargin=*, align=left]
  \item \textit{Fix the shape of PReLU weight}
  \item \textit{Add self.isCompleteTensor()}
\end{itemize}  &
\textbf{Commit Messages:} 
\begin{itemize}[leftmargin=*, align=left]
  \item \textit{new utility method}
  \item \textit{added \texttt{<?>} removed unused imports}
  \item \textit{better javadoc}
  \item \textit{Merge branch 'master' of \url{https://github.com/DeepSnowNeeL/javaparser.git}}
  \item \textit{another helper method and removed a useless check}
  \item \textit{helper method to add body to an ObjectCreationExpr}
  \item \textit{removed raw type warning}
  \item \textit{Fix for Type\texttt{<?>}}
  \item \textit{meh missed this one}
  \item \textit{more raw types fix + ObjectCreationExpr uses NodeWithType}
  \item \textit{fix test}
\end{itemize} &
\textbf{Commit Messages:} 
\begin{itemize}[leftmargin=*, align=left]
  \item \textit{LibLaserCut}
  \item \textit{first draft}
\end{itemize} \\
\hline
\end{tabularx}%
}
\label{table:samples}
\end{table*}

\section{Motivation} \label{background-sec}

In a well-crafted PR, the description should effectively summarize the main changes, provide the rationale behind them, and may also add context, such as the motivation for why the changes were made. This allows mergers to quickly and accurately identify the key modifications and make informed decisions without having to dive deeply into commit messages or \texttt{diff} files. Despite several description generation models being proposed to automatically generate the descriptions \cite{liu2019automatic, fang2022prhan, kuang2021automatic}, these models are often trained on possibly noisy PR description datasets, which raises questions about their performance.

For instance, \textbf{Example 1} in \Cref{table:samples} illustrates a partially noisy PR sample, with some of its commit messages lacking useful information. One of the commit messages used in the input sequence, ``\textit{Merge branch 'master' into fix-cme-on-start-passive-scan}'', does not offer any significant insights into the changes. We refer to such commit messages as \textbf{trivial commit messages}, following Liu et al. \cite{liu2018neural}. Consequently, since trivial commit messages lack contextual information about the PR, a model trained on such samples would struggle to generate a descriptive description similar to that in \textbf{Example 1}. It would be more beneficial if the commit messages aligned with the PR descriptions and offered a verbal explanation of the content or rationale behind the changes.

Similar to trivial commit messages, the existence of \textbf{trivial PR descriptions} further complicates the understanding of the underlying changes. These auto-generated or hastily written descriptions often fail to provide informative context about the updates. For example, the original PR description of \textbf{Example 2} in \Cref{table:samples}, \textit{fix issue \#21271}, does not offer any substantial information about the nature of the issue or the implemented solution. A high-quality alternative description would instead explain that the fix is related to the shape of the PReLU weight, aligning with the commit messages. As a result, trivial PR descriptions diminish the clarity of the changes, making it difficult to understand the PR's nature.

On the other hand, \textbf{Example 3} in \Cref{table:samples} demonstrates an original PR description that fails to adequately describe the PR due to its irrelevant context. Besides acknowledging an issue with formatting, the PR author did not provide any information about the changes made to the source code. Consequently, the current PR description does not effectively summarize the PR. Descriptions that include subjective messages or excessive details about implementation and test results often fail to capture the main changes.

Finally, \textbf{Example 4} in \Cref{table:samples} presents a situation where the PR description contains more than twice as many words as the commit messages. However, since PR description generation is treated as a text summarization problem, where the summary is derived from the commit messages, it is generally expected that the word count of a PR's commit messages should be sufficient to ensure the generated description closely matches the ground truth. Consequently, the commit messages of such PRs might be too short to provide sufficient information to generate a detailed description successfully.

These four examples highlight the presence of low-quality PRs that were used to train automatic description models. The existence of such noisy PRs motivates us to propose new cleaning heuristics to ensure that automatic generation models can be trained on a high-quality dataset, resulting in more reliable descriptions. Then, high-quality descriptions can help save time for developers and improve the reviewing process.

\section{Methodology} \label{meth-sec}

In this section, we explain the methodology of this study. In \Cref{cleaning-sec}, we define our four cleaning heuristics to eliminate noisy PR samples and improve the quality of the input sequence. Then, we introduce the text summarization models used for automatic evaluation in \Cref{models-sec}.

\subsection{Proposed Cleaning Heuristics} \label{cleaning-sec}

In order to eliminate noise in PR description datasets and build a suitable sample set, we propose the following four cleaning heuristics. 

\textbf{\textit{Heuristic Rule 1:}} \textit{Trivial commit messages that follow a specific pattern and contain little useful information will be filtered out from the input sequence.} 

The presence of trivial commit messages in the input sequence leads to the generation of uninformative descriptions. These commit messages are often of questionable quality as they are automatically generated and provide minimal information. The specific commit message patterns that will be discarded from the dataset are detailed in \Cref{commit-patterns}, which combines findings from the works of \cite{liu2018neural}, \cite{zhang2022automatic}, and our own observations. We use this heuristic to filter commit messages similar to \textit{Example 1} in \Cref{table:samples} from the input sequence. \\

\begin{table}[h!]
  \begin{center}
    \caption{Trivial Commit Message Patterns}
    \label{commit-patterns}
    \begin{tabular}{l|l} 
      \textbf{Starts with} & \textbf{Possible Continuation} \\\hline
      merge & .*? branch .*? into / branch $\backslash$\lq/ pull request \#$\backslash$d+ \cite{zhang2022automatic}\\ 
      update & changelog / gitignore / readme \cite{liu2018neural}\\
      modify & makefile \cite{liu2018neural}\\
      add & gitignore \cite{liu2018neural}\\
      closes & \#$\backslash$d+
    \end{tabular}
  \end{center}
\end{table}

\textbf{\textit{Heuristic Rule 2:}} \textit{Trivial PR descriptions that contain no useful information about the changes will be removed from the dataset.} 

Similar to trivial commit messages, trivial PR descriptions are those that can be written with minimal effort and fail to summarize the PR. As we want the models to generate non-trivial PR descriptions, PRs with trivial descriptions will be discarded. By examining our dataset and the patterns for trivial PR titles identified by Zhang et al. \cite{zhang2022automatic}, we have defined the criteria for trivial PR descriptions in \Cref{description-patterns}. We use this heuristic to filter PR samples with descriptions similar to \textit{Example 2} in \Cref{table:samples}.

\begin{table}[h!]
  \begin{center}
    \caption{Trivial PR Description Patterns}
    \label{description-patterns}
    \begin{tabular}{l|l} 
      \textbf{Starts with} & \textbf{Possible Continuation} \\\hline
      rolling & up .* /down .* \cite{zhang2022automatic}\\ 
      roll & engine .* /plugins .* \cite{zhang2022automatic}\\
      merge to & .* \cite{zhang2022automatic}\\
      revert & .* \cite{zhang2022automatic}\\
      update & changelog/gitignore/readme/current master \cite{zhang2022automatic}\\
      fix & issue \#$\backslash$d+\\
    \end{tabular}
  \end{center}
\end{table}

\textbf{\textit{Heuristic Rule 3:}} \textit{PRs with descriptions missing over 80\% of the words from the input sequence will be considered as \textbf{irrelevant} and will be removed from the dataset.} 

In practice, a summarization typically includes some of the original words from the input sequence. Therefore, if a PR description has a significant portion of its words (e.g., \textit{x\%}) missing in the input, it is unlikely to effectively summarize the PR. We apply this heuristic to filter out samples like \textit{Example 3} in \Cref{table:samples}, setting \textit{x} to 80 in our case following \cite{zhang2022automatic} for the choice of the threshold value. \\

\textbf{\textit{Heuristic Rule 4:}} \textit{PRs whose input sequence contains half or fewer words than the ground truth description will be considered as \textbf{inadequate} and will be removed from the dataset.}

As the task of description generation is formulated as a text summarization problem, the target sequence is designed to contain fewer words than the input sequence. Consequently, for effective training, a good PR sample should ideally have more words in the input sequence than in the ground truth description. This step is critical because when the input sequence contains significantly fewer words compared to the ground truth description, the generated description may differ substantially from its ground truth. This discrepancy can potentially lead to a negative impact on the training process, resulting in models that generate very short descriptions compared to the ground truth.

Supporting this perspective, Fang et al. \cite{fang2022prhan} highlight that the PR descriptions generated by PRSummarizer \cite{liu2019automatic}, one of the four models we employ, consistently exhibit shorter lengths than the ground truth PR descriptions, with 67.2\% of them being at least 1.5 times shorter. Therefore, to address this discrepancy, we establish a heuristic by setting a ratio threshold (denoted as $x$) between the number of words in the input sequence and the ground truth description. If the ratio is lower than $x$, we remove the PR from the dataset. After observing the initial uncleaned dataset, we set $x$ as 1/2. We apply this final heuristic to filter out samples like \textit{Example 4} in \Cref{table:samples}.

\subsection{Summarization Models} \label{models-sec}

In this section, we provide details about the four abstractive summarization models used in our experiments. We chose to employ abstractive models because they can generate target sentences different from the original sentences in the input sequence \cite{el2021automatic}.

The first two models, PRSummarizer \cite{liu2019automatic} and iTAPE \cite{chen2020stay}, are domain-specific summarization models. PRSummarizer aims to create descriptions of PRs from commit messages and source code comments, while iTAPE focuses on generating automatic issue titles. However, both of these models must be trained from scratch, making them heavily reliant on the quality of the dataset used for training. BART \cite{lewis2019bart} and T5 \cite{raffel2019exploring}, on the other hand, are general-purpose text summarization tools. Both are based on the Transformer \cite{vaswani2017attention} architecture and have been pre-trained on large datasets. This pre-training makes them less dependent on the training set we employ, and they tend to perform better compared to domain-specific summarization models such as iTAPE and PRSummarizer \cite{zhang2022automatic}.

\section{Evaluation} \label{eval-sec}

This section provides an overview of the automatic and manual evaluation metrics employed in our study. We use the ROUGE metrics, which are used to analyze the performance of the summarization models, in \Cref{metric-sec}. Subsequently, we explain the details of two setups designed to assess the impact of our heuristics from a human perspective in \Cref{manuel-sec}, representing the focal point of the manual evaluation.

\subsection{Automatic Evaluation} \label{metric-sec}

We evaluate our approach with Recall-Oriented Understudy for Gisting Evaluation (ROUGE) \cite{lin2004rouge} metric following the works of \cite{liu2019automatic, fang2022prhan, kuang2021automatic, zhang2022automatic}. We mainly focus on ROUGE-N (N=1,2) and ROUGE-L metrics. For each metric, we calculate the Recall, Precision, and F1 score. The formulas for these are shown as follows:

\[
R_{\textit{rouge-n}} = \frac{\sum_{(gen, ref) \in S} \sum_{\textit{gram}_n \in \textit{ref}} \textit{Cnt}_{\textit{gen}}(\textit{gram}_n)}{\sum_{(gen, ref) \in S} \sum_{\textit{gram}_n \in \textit{ref}} \textit{Cnt}_{\textit{ref}}(\textit{gram}_n)}
\]

\[
P_{\textit{rouge-n}} = \frac{\sum_{(gen, ref) \in S} \sum_{\textit{gram}_n \in \textit{ref}} \textit{Cnt}_{\textit{gen}}(\textit{gram}_n)}{\sum_{(gen, ref) \in S} \sum_{\textit{gram}_n \in \textit{gen}} \textit{Cnt}_{\textit{gen}}(\textit{gram}_n)}
\]

\[
F1_{\textit{rouge-n}} = \frac{2 \cdot R_{\textit{rouge-n}} \cdot P_{\textit{rouge-n}}}{R_{\textit{rouge-n}} + P_{\textit{rouge-n}}}
\]

where \textit{gen}, \textit{ref}, and \textit{S} denote the PR description generated by the models, reference description, and test dataset, respectively. $\textit{gram}_n$ is the n-gram phrase whereas $\textit{Cnt}_{\textit{\{gen, ref\}}}(\textit{gram}_n)$ denote the number of occurrences of $\textit{gram}_n$ in $\textit{gen}$ and $\textit{ref}$, respectively. To summarize, $R_{\textit{rouge-n}}$ indicates the percentage of n-grams in the reference description covered by the generated description, while $P_{\textit{rouge-n}}$ measures the percentage of ``correct'' n-grams in the generated descriptions. In this context, ``correct'' n-grams refer to relevant n-grams appearing in reference descriptions. $F1_{\textit{rouge-n}}$ is the measure that combines both precision and recall. The formulas for ROUGE-L are similar to those for ROUGE-N. However, instead of n-grams, ROUGE-L measures the longest common subsequence between the reference description and the generated description \cite{lin2004rouge}. Following Zhang et al. \cite{zhang2022automatic}, we use the metric present in the Hugging Face datasets library \cite{lhoest2021datasets}.

\subsection{Manual Evaluation} \label{manuel-sec}

\begin{figure*}[ht!]
    \small
    \centering
    \includegraphics[width=0.76\textwidth]{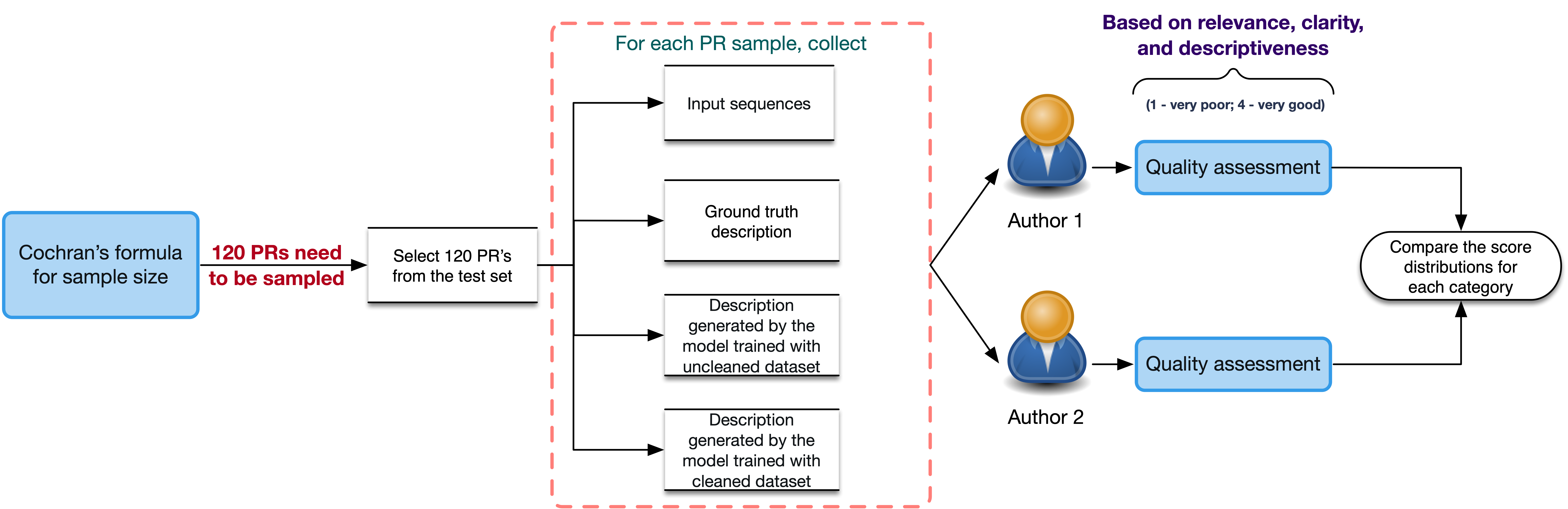}
    \caption{Overview of the First Stage of Manual Evaluation}
    \label{fig:first-manual-assessment}
\end{figure*}

\begin{figure*}[h!]
    \centering
    \includegraphics[width=0.76\textwidth]{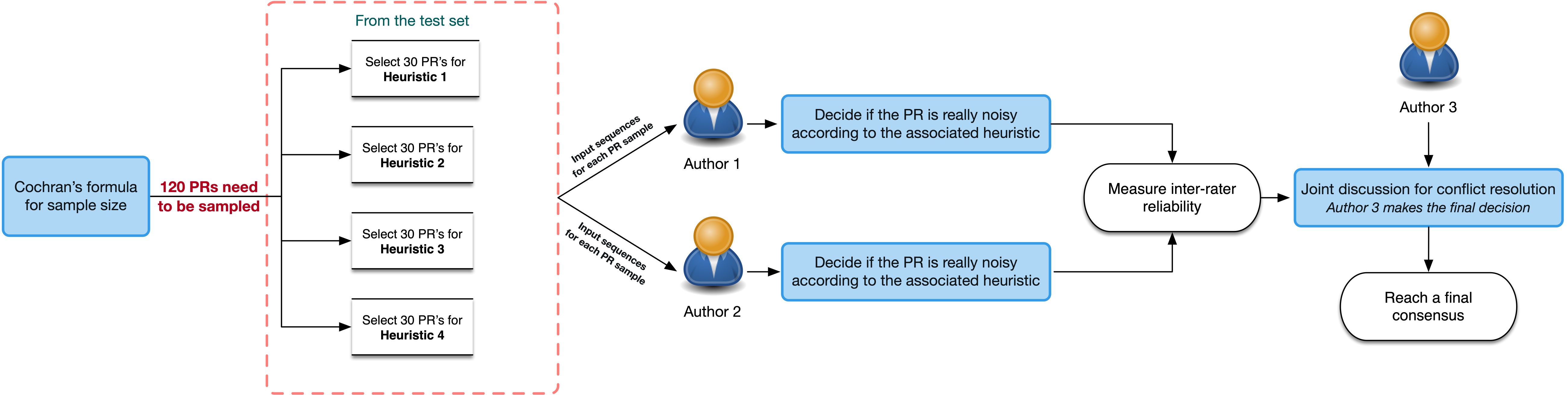}  
    \caption{Overview of the Second Stage of Manual Evaluation}
    \label{fig:second-manual-assessment}
\end{figure*}

In addition to automatically evaluating the generated descriptions, we also conducted a manual evaluation to gain a better understanding of the effectiveness of our cleaning heuristics and the quality of descriptions produced by the models when trained on a cleaned dataset.
As pointed out by Zhang et al. \cite{zhang2022automatic}, due to the reliance on n-gram overlap in calculating ROUGE scores, the generated summaries may exhibit semantic inaccuracies, even in cases where a notably high ROUGE score is achieved. This occurs because ROUGE scores solely assess the similarity of words between sentences and, as a result, cannot assess the comprehensibility of the generated summaries by the models \cite{van2021human}.  

In the first stage of our manual evaluation, shown in \Cref{fig:first-manual-assessment}, two authors independently evaluated the qualities of generated descriptions produced by BART, the best-performing approach, when it was trained with an uncleaned dataset versus a cleaned dataset. To facilitate the comparison of the two sets of descriptions, 120 PRs from the cleaned test set were sampled along with their generated descriptions. The sampling was conducted with a 92\% confidence level and an 8\% margin of error, using Cochran’s sample size formula \cite{cochran2007sampling}, following the approach of Zhang et al. \cite{zhang2022automatic} for the choice of values. For each sample, the two authors received an input sequence consisting of commit messages and the ground truth description, along with two generated descriptions: one obtained from BART when it was trained on the cleaned dataset, and the other when it was trained on the uncleaned dataset. Importantly, the authors were blinded to the model and the training regime, meaning they did not know which description came from the cleaned or uncleaned dataset, thus eliminating any potential bias in their assessments. Then, the authors were tasked with reviewing the input sequence and the two descriptions, assigning scores (1 for very poor, 2 for poor, 3 for good, and 4 for very good) to each description based on the following criteria:
\begin{itemize}
    \item \textbf{Relevance:} How closely does the generated description align with the commit messages and context provided in the PR?
    \item \textbf{Descriptiveness:} How well does the generated description convey the intended information and summarize the PR?
    \item \textbf{Clarity:} How easily can the reader understand the generated description, and how concise is the description?
\end{itemize}

As the final step, the distribution of scores for each category was compared for the two sets of descriptions.

The second stage, shown in \Cref{fig:second-manual-assessment}, aimed to evaluate the effectiveness of each heuristic independently and determine if they truly identify and eliminate noisy PR samples. To achieve this, 30 PRs were chosen for each heuristic out of the 452,348 PRs identified as noisy. In total, 120 PRs were sampled, taking into account the existence of four distinct heuristics. Cochran’s sample size formula \cite{cochran2007sampling} was once again employed for sampling, ensuring a 92\% confidence level and 8\% margin of error. In this phase, the authors received the input sequence and ground truth description of the PR samples eliminated by each heuristic. Their objective was to assess the accuracy of each heuristic in identifying these PRs as noisy.
Each PR sample was labeled as either a \textit{true positive}, indicating that the PR was correctly identified as noisy, or a \textit{false positive}, indicating that the PR was not noisy despite being flagged by the heuristic.

Following the labeling of each PR, we computed inter-rater reliability using Cohen’s kappa score. This metric serves as a statistical measure of agreement between the two authors \cite{Kappa}. Discrepancies in the labels provided by the two authors were addressed in a subsequent session that included the participation of a third author. The objective of this meeting was to resolve disagreements through discussion and consensus-building. With these finalized labels, we gain valuable insights into the accuracy of each heuristic. 

\section{Experimental Setup} \label{setup-sec}

\subsection{Data Collection}

Even though Liu et al. \cite{liu2019automatic} shared a dataset for PR description generation, we observed that the number of PRs they included (41K+) may not be enough to test the efficiency of our heuristics. Furthermore, as pointed out by Zhang et al. \cite{zhang2022automatic}, their dataset only included PRs from Java projects, which is not diverse. To overcome this, we collected PRs from the following six repository lists in GitHub: Top-100 most-starred, Top-100 most-forked, and the Top-100 most-starred repositories of the following languages: Python, Java, JavaScript, and C++. The choice of programming languages was kept the same as in the work of Zhang et al. \cite{zhang2022automatic}, as these languages are among the most commonly used. Given that the six repository lists contain intersecting repositories and after removing 25 empty repositories, we crawled 513 repositories in total. In this process, we used the GitHub GraphQL API \footnote{\url{https://docs.github.com/en/graphql}} to crawl the PRs and fetch the commit messages and descriptions of each PR. In the end, we obtained 2,185,382 PRs.

\subsection{Data Preprocessing}

\begin{table*}[ht!]
  \centering
  \caption{Preprocess Statistics of the Collected PRs}
  \label{tab:preprocess}
  \resizebox{\linewidth}{!}{\begin{tabular}{llccccc}
    \toprule
    \textbf{Initial PRs} & \textbf{$<$2 Commits} & \textbf{$>$20 Commits} & \textbf{Contains Non-ASCII Characters} & \textbf{Bot-written} & \textbf{Has Empty Description} & \textbf{Left PRs}\\
    \midrule
    2,185,382 & 1,319,061 & 62,983 & 76,931 & 5329 & 95,260 & 625,818\\
    \bottomrule
  \end{tabular}}
\end{table*}

Before applying our cleaning heuristics to the dataset, we perform data preprocessing on the collected PRs. We begin by excluding PRs with fewer than two commits or more than 20 commits, following the approach of Liu et al. \cite{liu2019automatic}. This decision was made because we can directly use the single commit message as the PR description when the number of commits is less than two, without needing to generate a description. Additionally, PRs with too many commits are typically used for synchronization purposes rather than describing a specific contribution made by a developer \cite{liu2019automatic}. 

We then filter out two categories of PRs: those containing non-ASCII characters and those authored by bots. To identify PRs authored by bots, we check if the author is labeled as a \texttt{bot} in the PR's author field, a feature available through the GitHub API. We remove bot-written PRs because they typically contain automatically generated descriptions that are simplistic and add little value to our analysis. Instead, we focus on PRs crafted by individuals. Finally, we remove the checklists in PR descriptions, following Liu et al. \cite{liu2019automatic} once again. These checklists are automatically generated and usually outline the standard steps required to finalize a PR \cite{liu2019automatic}, rather than focusing on the content of the PR. After our data preprocessing, we have 625,818 PRs left in the dataset from the initial 2,185,382. \Cref{tab:preprocess} shows the preprocess statistics of the collected PRs. We split our dataset into training, validation, and test sets with a ratio of 8:1:1. In the end, we obtained 500,655 PRs for training, 62,581 PRs for validation, and 62,582 PRs for testing. 


\subsection{Cleaning of the Dataset}

\begin{figure}[!htbp]
    \centerline{{\includegraphics[width=0.85\columnwidth]{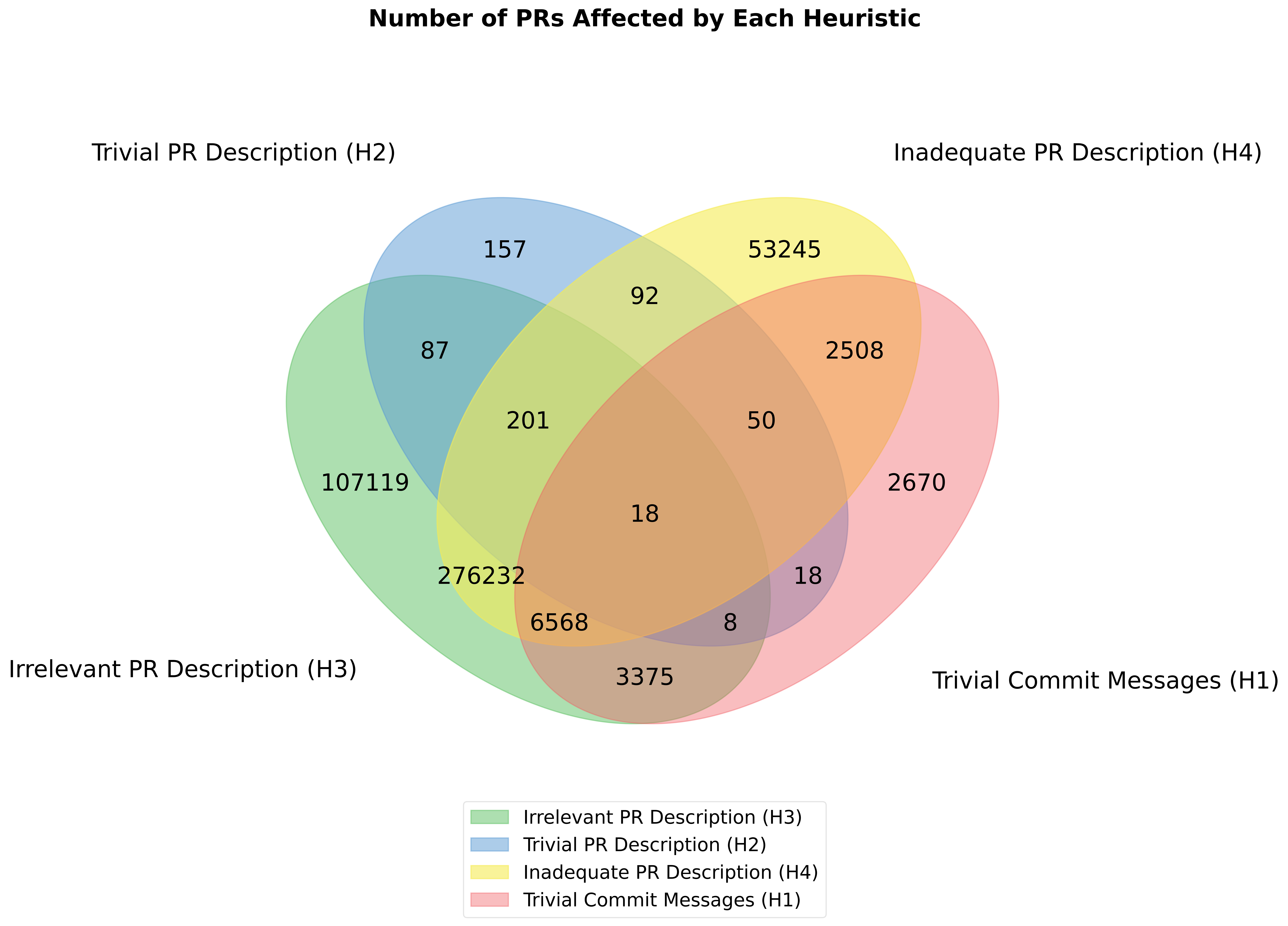}}}
    \caption{Number of PRs Affected by Each Heuristic}
    \label{normal-cleaned-fig}
\end{figure}

To assess the performance of the models, we applied our cleaning heuristics to the training, validation, and test sets individually. After applying our heuristics, the size of our training set was reduced from 500,655 PRs to 138,794 PRs, the validation set to 17,110 PRs, and the test set to 17,566 PRs. To maintain the ratio between the training set and test and validation sets, we sampled 15,000 PRs from the cleaned test set and 15,000 PRs from the cleaned validation set to form the final test and validation sets. This final test set is used to evaluate the performance of the models trained on the cleaned training set and the uncleaned training set separately.
After forming our sets, we calculated the number of PRs affected by each heuristic. We represented the numbers in a Venn diagram in \Cref{normal-cleaned-fig}, acknowledging that a PR can be affected by multiple heuristics. The numbers reflect the total PRs affected by each heuristic across the training, validation, and test sets.

We observed that the majority of the cleaning is attributed to \textbf{heuristic rule 4}, which eliminates PRs similar to \textit{Example 4} in \Cref{table:samples}. This means that approximately 54\% of the PRs contained input sequences shorter than half the length of the ground truth description. The decision to include the $1/2$ factor was intentional to preserve some PRs potentially beneficial for training, yet a general discrepancy persists between the lengths of the input sequences and the ground truth descriptions. In total, 452,348 PRs were affected by our four heuristics which corresponds to 72.3\% of the PRs obtained after preprocessing. 

\section{Results} \label{results-sec}

\subsection{\textit{RQ1: Automatic Evaluation of Cleaning Heuristics}}

To investigate the effect of our cleaning heuristics, we analyzed the performance of four models trained with cleaned and uncleaned datasets separately using the ROUGE-1, ROUGE-2, and ROUGE-L metrics. For a fair comparison, the models were tested on the same cleaned test set, as mentioned earlier. We selected the cleaned test set to ensure that the evaluation was conducted in a noise-free environment, allowing for a more accurate comparison. We then reported the average scores of the models trained on the uncleaned and cleaned datasets.  The results can be seen in \Cref{tab:summarization_metrics}.

For the PR description generation task, the application of our cleaning heuristics increases the average F1 scores for ROUGE-1, ROUGE-2, and ROUGE-L metrics, as seen in \Cref{tab:summarization_metrics}. The average F1 scores of the four models increase by 8.6\%, 8.7\%, and 8.5\% for ROUGE-1, ROUGE-2, and ROUGE-L, respectively. Additionally, both average precision and recall scores are higher for each ROUGE metric for the models trained with the cleaned dataset. Higher precision in this context means that the generated descriptions after applying our heuristics contain more relevant and accurate content, reducing the inclusion of unnecessary or out-of-context information. These results indicate that, compared to models trained on the initial uncleaned dataset, the models trained with the cleaned dataset can summarize a PR more precisely as they perform better on the same test set.

\begin{table*}[t!]
  \centering
  \caption{Average Comparative Performance Metrics for Models on Uncleaned vs. Cleaned Datasets}
  \label{tab:summarization_metrics}
  \resizebox{0.95\textwidth}{!}{\begin{tabular*}{\textwidth}{l@{\extracolsep{\fill}}ccccccccc}
    \cmidrule[\heavyrulewidth](l{3pt}r{3pt}){1-10}
    \textbf{Dataset} & \multicolumn{3}{c}{\textbf{ROUGE-1}} & \multicolumn{3}{c}{\textbf{ROUGE-2}} & \multicolumn{3}{c}{\textbf{ROUGE-L}} \\
    \cmidrule(l{3pt}r{3pt}){2-4} \cmidrule(l{3pt}r{3pt}){5-7} \cmidrule(l{3pt}r{3pt}){8-10}
    & \textbf{Precision} & \textbf{Recall} & \textbf{F1} & \textbf{Precision} & \textbf{Recall} & \textbf{F1} & \textbf{Precision} & \textbf{Recall} & \textbf{F1} \\
    \cmidrule(l{3pt}r{3pt}){1-10}
    \textbf{Uncleaned} & 64.15 & 48.10 & 53.44 & 52.83 & 39.23 & 44.73 & 60.60 & 45.37 & 50.83 \\
    \cmidrule(l{3pt}r{3pt}){1-10} 
    \textbf{Cleaned} & 68.50 & 51.64 & 58.03 & 56.32 & 42.91 & 48.63 & 64.41 & 48.95 & 55.17 \\
    \cmidrule(l{3pt}r{3pt}){1-10}
    \textit{Improvement} & \textit{+6.78\%} & \textit{+7.36\%} & \textit{+8.60\%} & \textit{+6.60\%} & \textit{+9.38\%} & \textit{+8.71\%} & \textit{+6.28\% }& \textit{+7.89\%} & \textit{+8.53\%} \\
    \cmidrule[\heavyrulewidth](l{3pt}r{3pt}){1-10}
  \end{tabular*}}
\end{table*}

\begin{figure*}[h!]
    \centering
    \includegraphics[width=\textwidth]{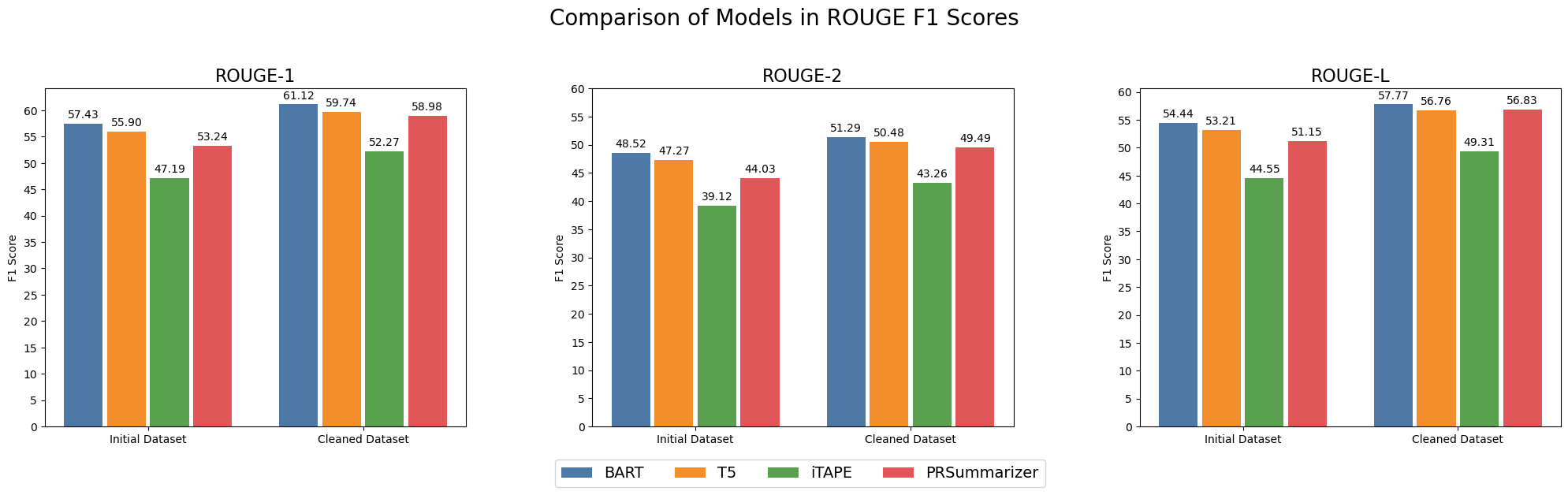} 
    \caption{F1 Scores of All Four Models for ROUGE-1, ROUGE-2, and ROUGE-L Metrics}
    \label{fig:f1-scores}
\end{figure*}

Among the four abstractive approaches, BART and T5 demonstrated better performance on both the initial uncleaned and cleaned datasets compared to the other two abstractive methods, as illustrated in \Cref{fig:f1-scores}. This result is not surprising, given that BART and T5 were both pre-trained on large general-purpose corpora. In contrast, iTAPE and PRSummarizer were trained only on our datasets. Even though all approaches have reasonably good F1 scores, when combined with the power of pre-training on large corpora and the application of our heuristics, BART achieves the highest overall F1 scores for each ROUGE metric, with 61.12 for ROUGE-1, 51.29 for ROUGE-2, and 57.77 for ROUGE-L. As a result, we believe that BART, when trained on a dataset cleaned by our heuristics, shows promising potential for use in automatic PR description generation tasks.

The fact that iTAPE and PRSummarizer are solely trained on our dataset provides a clearer understanding of the effectiveness of our cleaning heuristics. By examining the F1 scores of PRSummarizer in \Cref{fig:f1-scores}, we observe that the F1 scores increase from 53.24 to 58.98 (a 10.78\% relative increase) for ROUGE-1, from 44.03 to 49.49 (a 12.40\% relative increase) for ROUGE-2, and from 51.15 to 56.83 (a 11.11\% relative increase) for ROUGE-L. This corresponds to an average increase of 11.43\% for all F1 scores. A similar calculation for iTAPE yields an average increase of 10.67\%. When compared to the average increases of the pre-trained models BART (6.07\%) and T5 (6.77\%), both PRSummarizer and iTAPE show better improvement when trained on the cleaned dataset. Therefore, this suggests that our cleaning heuristics are particularly effective 
when a text summarization model needs to be trained from scratch.

\subsection{\textit{RQ2: Manual Evaluation of Cleaning Heuristics}}

\begin{figure}[htbp]
    \centering
    \includegraphics[width=1.05\linewidth]{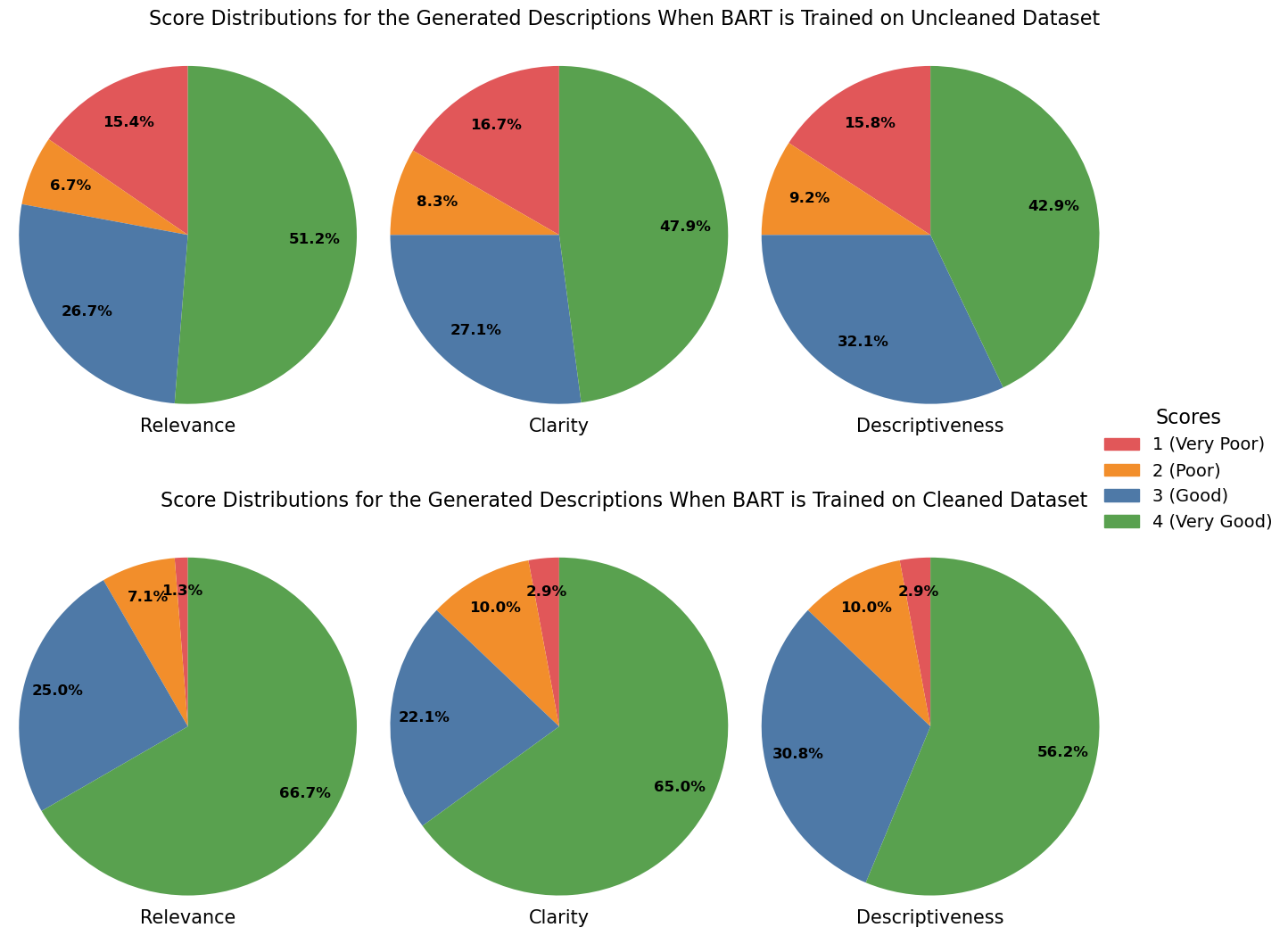}
    \caption{Score Distributions of Generated Descriptions for Each Category}
    \label{fig:author-labels}
\end{figure}

As explained earlier, our manual evaluation consists of two stages. For the first stage, \Cref{fig:author-labels} shows the score distributions of PR descriptions generated by BART, evaluated on three aspects by two authors. These descriptions were chosen from BART as it was the best-performing model in terms of F1 scores. It is clear that the descriptions generated by BART when trained on the cleaned dataset are better in all three aspects: \textit{relevance}, \textit{clarity}, \textit{descriptiveness}. 
Among the three aspects, the lowest percentage of very poor descriptions, with a score of 1, belongs to \textit{relevance}, at 1.3\%. Consequently, this reduction in the percentage of very poor descriptions results in a 15.5\% increase in the highest quality descriptions, with a score of 4, as can be seen in \Cref{fig:author-labels}.

This increase can primarily be attributed to \textit{heuristic rule 3}, which was responsible for eliminating \textbf{irrelevant} PR samples. For example, a generated PR description with better quality is given in the first column of Example 1 in \Cref{tab:description_comparison} when BART was trained with the cleaned dataset. The description next to it, however, was generated by BART trained on the uncleaned dataset. Despite using the same input sequence in both cases, BART trained on the uncleaned dataset fails to distinguish irrelevant contexts, including PR template words like ``\texttt{please fill in this template}'' or ``\texttt{use a meaningful title for the pull request}''. Consequently, it generates a description that fails to explain any of the changes suggested in the PR. In contrast, the description generated by BART when trained on the cleaned dataset is significantly more relevant to the PR. The presence of such examples explains the 15.5\% increase in the distribution of very good quality descriptions in the relevance category when BART was trained on the clean dataset.

\begin{table*}[htbp]
\small
\centering
\caption{Sample Descriptions from BART Trained on Cleaned vs. Uncleaned Dataset}
\label{tab:description_comparison}
\resizebox{0.95\linewidth}{!}{\begin{threeparttable}\begin{tabular}{@{}p{0.11\textwidth}p{0.44\textwidth}p{0.44\textwidth}@{}}
\toprule
 & \textbf{Descriptions Generated by BART Trained on the Uncleaned Dataset} & \textbf{Descriptions Generated by BART Trained on the Cleaned Dataset} \\ \midrule
Example 1\tnote{a} &  \textbf{Generated Description:} please fill in this template. use a meaningful title for the pull request. include the name of the package modified. test the change in your own code. (compile and run.) add or edit tests to reflect the change. (run with npm test.) follow the advice from the readme. avoid common mistakes. run npm run lint package-name (or tsc if no tslint.json is present). select one of these and delete the others:  & \textbf{Generated Description:} added optional type information for the sequelize's import define function
 \\
Example 2\tnote{b} & \textbf{Generated Description:} fixes: \#13374 
& \textbf{Generated Description:} this patch creates rbdmap shell script that is called from init-rbdmap init script. the patch also renames src/rfmap configuration file to src/etc-rfmap so that it can be installed via build system directly. finally, the patch accommodates these changes in spec file and build system. fixes: \#13374
\\
\bottomrule
\end{tabular}
  \begin{tablenotes}
  \item[a] \url{https://github.com/DefinitelyTyped/DefinitelyTyped/pull/6590}
  \item[b] \url{https://github.com/ceph/ceph/pull/6479}
  \end{tablenotes}
  \end{threeparttable}}
\end{table*}

Another increase in the percentage of very good quality descriptions with a score of 4 is observed in the \textit{descriptiveness} category, with an average increase of 13.3\%. We believe that the combination of \textit{heuristic rule 1} and \textit{heuristic rule 2} was particularly effective in this regard, as it enabled us to obtain a dataset filtered of trivial commit messages and trivial PR descriptions, as showcased in \textit{Example 1} and \textit{Example 2} in \Cref{table:samples}, respectively. When BART was trained on the cleaned dataset, the generated description was much more successful in detailing the changes made. However, the description obtained from BART trained on the uncleaned dataset, \texttt{fixes \#13374}, does not provide any detail. Even though such descriptions can be useful for reviewers to directly navigate to the issue that the PR addresses, we believe that using these descriptions in training can negatively impact the models' ability to generate more descriptive content capable of summarizing the PR. 

Finally, we observe an average increase of 17.1\% in the \textit{clarity} category in the distribution of very good quality descriptions. The main reason for this improvement is that the descriptions generated from BART trained on the cleaned dataset feature more structured sentences that are easier to understand. Furthermore, the absence of specific template words in the descriptions makes the generated descriptions clearer and easier to comprehend. Hence, the first stage of the manual evaluation demonstrates that descriptions obtained from BART are more descriptive, easier to understand, and more relevant to the changes proposed in the PR when BART is trained on the cleaned dataset.

\begin{table}[htbp]
  \centering
  \caption{Performance Metrics For Different Heuristics}
  \label{tab:heuristic_accurcay}
  \begin{threeparttable}
    \begin{tabular}{lcccc}
      \toprule
      Heuristic & TP & FP & Accuracy & Inter-rater Reliability \\
      \midrule
      \textit{Heuristic Rule 1} & 27 & 3 & 90.00\% & 0.71 \\
      \textit{Heuristic Rule 2} & 20 & 10 & 66.67\% & 0.92 \\
      \textit{Heuristic Rule 3} & 18 & 12 & 60.00\% & 0.86  \\
      \textit{Heuristic Rule 4} & 25 & 5 & 83.34\% & 0.87 \\
      \bottomrule
    \end{tabular}
    \begin{tablenotes}
      \item[] TP indicates ``True Positive'' and FP indicates ``False Positive''
    \end{tablenotes}
  \end{threeparttable}
\end{table}

In the second stage of the manual evaluation, we examined the individual PR samples identified as noisy by each heuristic and computed the accuracy of the heuristics. \Cref{tab:heuristic_accurcay} displays the accuracy and the inter-rater reliability scores of each heuristic. \textit{Heuristic rule 1} achieved an accuracy of 90\% among the 30 examined PR samples. This result is not surprising, as the trivial commit messages removed by this heuristic seemed to lack a contribution to the generation of high-quality descriptions. The inter-rater reliability was found to be 0.71, indicating moderate agreement. For \textit{heuristic rule 2}, the inter-rater reliability score was 0.91, indicating almost perfect agreement among the authors, while the accuracy was found to be 66.67\%. Upon closer inspection of the 10 false positive samples marked as noisy by this heuristic, it was discovered that five of them were eliminated due to containing the \texttt{fix issue \#d+} pattern. We observed that although descriptions containing only this pattern can be considered noisy, developers sometimes provide additional information alongside the fixed issue number, which can be valuable while training the models. Therefore, such cases were marked as false positives by the authors.

The inter-rater reliability of \textit{heuristic rule 3} is found as 0.86, which shows strong agreement. However, the accuracy of this heuristic was lower compared to the others at 60\%. To understand the main reason, we investigated each sample marked as false positive and discovered that some ground truth descriptions with more than 80\% of words missing from the input sequence could still be useful for training the models. For example, the description \texttt{``redoes a phase of type checking to validate patterns for minimality and completeness. Cherry-picked from \#8908, \#9121, \#9155, and \#9180.''} was eliminated because the mentioned numbers were not contained in the input sequence. However, the first sentence should not be considered a noisy description, as it successfully explains the changes proposed by the PR. Such examples were ultimately marked as false positives by the authors. On the other hand, the true positive samples were primarily attributed to containing subjective sentences or, more commonly, the presence of PR template words. 

Lastly, \textit{heuristic rule 4} achieved an inter-rater reliability score of 0.87 and an accuracy of 83.34\%. The high accuracy of this heuristic was expected because samples containing half or fewer words in the input sequence compared to the ground truth description were ultimately classified as noisy, as they could not generate high-quality descriptions with so few words. 
While a larger-scale study may be required, our manual evaluation provides preliminary evidence that our heuristics effectively detect noisy PR samples. Still, we acknowledge that the performance of the heuristics might vary with different threshold values for \textit{heuristic rule 3} and \textit{heuristic rule 4}, and different patterns for \textit{heuristic rule 1} and \textit{heuristic rule 2}.

\section{Threats to Validity} \label{validity-sec}

One threat to validity involves the specific threshold values employed in \textit{heuristic rules 3} and \textit{4}. In \textit{heuristic rule 3}, PR samples are discarded if their descriptions have over 80\% of words missing in the input sequence, as it is assumed that such PRs contain out-of-context ground truth descriptions, making them unsuitable for training. While this heuristic effectively removes non-summary PR descriptions, it is possible that some PR samples that could have been useful for training were also eliminated. Similarly, \textit{heuristic rule 4} eliminates nearly half of the data due to a length mismatch between the input sequence and the ground truth description. The eliminated PR descriptions could have also been of high quality and could have been used in training. We acknowledge that using different threshold values can result in a training set with better or worse PR samples and thus affect the performance of the models. 

One notable threat to validity concerns our experimental setup. Following \cite{zhang2022automatic}, we selected four state-of-the-art summarization methods to automatically evaluate our approach. Two of these models, iTAPE \cite{chen2020stay} and PRSummarizer \cite{liu2019automatic}, were domain-specific and trained from scratch, whereas the other two, BART \cite{lewis2019bart} and T5 \cite{raffel2019exploring}, were general-purpose methods. However, our results may vary when different models are employed. Similar to the choice of models, the selection of datasets also presents a threat to validity. For instance, our dataset comprises only examples from public open-source PRs, which could affect the generalizability of the results to private or commercial settings. Additionally, we utilized datasets spanning four programming languages, which may introduce biases depending on the language-specific characteristics and usage patterns. A potential future direction is to evaluate our heuristics with different summarization models, such as BERTSumExt \cite{liu2019text}, and to experiment with a broader range of programming languages.

Another threat to validity arises from our manual evaluation process. Like many other manual evaluations \cite{liu2019automatic, zhang2022automatic, chen2020stay}, our experimental results may be biased, and we cannot guarantee that each score assigned to every PR description is fair. However, to mitigate this threat, we calculated and presented the inter-rater reliability of the two authors using Cohen's kappa score. Furthermore, to solve the discrepancies during the labeling of each PR in the second stage, a session with the participation of a third author was held and the assessment was finalized accordingly.  

\section{Discussion} \label{disc-sec}
In this section, we provide a detailed exploration of how our cleaning heuristics translate into practical implications for software engineering, offering actionable insights and future work guidance for both practitioners and researchers.

\subsection{Implications for Practitioners}
Our approach to dataset cleaning and augmentation carries several implications for practitioners. Firstly, description or title generation tools in software engineering trained after applying our heuristics can demonstrate improved performance, as indicated by our results. One example of such a tool is AUTOPRTITLE \cite{irsan2022autoprtitle}, which is used for PR title generation and relies on various aspects of a PR, such as commit messages, as input. Applying our heuristics to the collected dataset can enhance the performance of these tools after they are trained with the cleaned dataset.

Another potential implication involves developing tools that monitor PR descriptions as they are composed. A tool using our heuristics might notify the developer if the content or length fails to satisfy the threshold our heuristics establish. Armed with this notification, the developer can revise the description to make it adequately detailed and thus ease the review process for potential PR reviewers.

One final implication is that by examining the PR descriptions classified as noisy by our heuristics, we gain insights into how PR descriptions should be written. This knowledge can be valuable for many practitioners in setting PR description guidelines. Future PRs could then follow these criteria to ensure higher quality and consistency.

\subsection{Implications for Researchers}

In addition to the implications for practitioners, our approach also holds significant implications for researchers. Firstly, we have constructed a benchmark dataset tailored for the task of automatic PR description generation. This dataset can serve as a valuable resource for further research on PR description generation models within the scope of software engineering. By eliminating noisy PR samples from this dataset, we ensure that newly developed models can be tested for performance without being significantly affected by noise. This provides researchers with an unbiased platform to evaluate their models. Moreover, if LLMs are to be used for the PR description generation task, they can be fine-tuned using datasets cleaned by our heuristics, thereby enhancing their task-specific performance and making them more effective in real-world applications.

Moreover, our work can be extended to private projects together with open-source projects. In private projects, the data and the content of the PRs may vary from topic to topic and thus require more extensive heuristics to eliminate noisy PR samples. This is due to the fact that PRs from private projects may be written in accordance with specific guidelines dictated by a company or an organization. Hence, a more detailed analysis of the heuristic rules may be required.

Our heuristics can also be tested with an input sequence that involves more information about the PRs, such as issue titles, code review comments, or code snippets from the \texttt{diff} files, to better mine the core idea of the PR and generate more accurate PR descriptions. The effect of our heuristics on such an enriched dataset remains a potential line of research.

Finally, our work highlights that prioritizing high-quality data not only improves model accuracy but also significantly enhances sustainability and efficiency in machine learning. Researchers should consider the environmental and cost benefits of focusing on high-quality data in model training, as this approach can lead to superior outcomes while mitigating the ecological footprint and lowering the costs associated with extensive model training.

\section{Related Work} \label{rel-work-sec}
In this section, we discuss the two research areas most relevant to our work: understanding pull requests in the context of software engineering and automating the documentation of software changes.

\subsection{Pull Requests in Software Engineering}

Pull-based software development has been increasingly studied in research mostly focusing on understanding the PRs empirically. For example, Gousios et al. \cite{gousios2015work, gousios2016work} conducted large-scale surveys to examine the work practices and challenges in pull-based development from the perspectives of mergers and contributors. They discovered that mergers face significant challenges in maintaining code and test quality, as well as prioritizing contributions due to the high volume of requests. Contributors, on the other hand, express concerns about being unaware of project status and encountering poor responsiveness from mergers.

Other works were concerned with PR-related tasks such as predicting the priority of PRs \cite{van2015automatically}, recommending reviewers for PRs \cite{yu2014reviewer}, or analyzing the factors affecting PR evaluation. For example, Rahman and Roy \cite{rahman2014insight} explored the impact of discussion texts, project-specific details, and information about developers, such as the experience of the developer on the acceptance rate of PRs. Later, Tsay et al. \cite{tsay2014influence} identified that the acceptance of PRs is influenced by both technical and social factors. Regarding reviewer recommendation, Yu et al. \cite{yu2014reviewer} suggested a technique to automatically recommend reviewers for a PR using its title, description, and developers' social relations. For PR prioritization, Veen et al. \cite{van2015automatically} proposed PRioritizer, a tool that uses machine learning techniques and various extracted features, such as the number of discussion comments, to prioritize PRs. 
As a result, all of this previous work has motivated us to automatically generate high-quality PR descriptions, which can benefit tasks that rely on PR descriptions.

\subsection{Automatic Documentation of Software Changes}

Software changes come in different levels of detail, such as commits, PRs, and releases. A commit represents a singular alteration made to a file or group of files within a repository. On the other hand, a PR encompasses the modifications made to a repository, typically comprising multiple commits. A release, meanwhile, refers to a version of the software that is deployable, packaged, and accessible for a broader audience to download and utilize. Some of the tools proposed to automatically generate commit messages include DELTADOC \cite{buse2010automatically}, CODISUM \cite{xu2019commit}, and ChangeScribe \cite{linares2015changescribe}. DELTADOC \cite{buse2010automatically} was able to summarize a commit by utilizing symbolic execution and path predicate analysis to generate the behavioral difference. Subsequently, by applying certain heuristic transformations, the commit message could be generated.
However, ChangeScribe \cite{linares2015changescribe} was the first tool to identify the stereotype of a commit from the abstract syntax trees and then generate a commit message using pre-defined filters and templates. CODISUM \cite{xu2019commit} on the other hand, was developed to address the limitations in previous commit message generation tasks: the oversight of code structure information and the challenge of handling out-of-vocabulary (OOV) issues. 
To mitigate the OOV issue, the tool adopts a copy mechanism, and for better code structure representation, joint modeling for both code structure and code semantics of the source code changes is used.
Later, Jiang et al. \cite{jiang2017automatically} developed an attentional encoder-decoder model to generate commit messages from \texttt{diffs}.

Researchers have also suggested various methods for automatically generating release notes \cite{moreno2014automatic, moreno2017arena, abebe2016empirical}. For instance, Moreno et al. \cite{moreno2014automatic, moreno2017arena} built a tool called ARENA that first summarizes each commit in a release and then uses manually defined templates to combine these summaries with their related information in the issue tracker to generate release notes. Abebe et al. \cite{abebe2016empirical} identified six types of information in release notes and used machine learning techniques to determine whether an issue should be included in release notes.

The initial endeavor of automatic PR description was undertaken by Liu et al. \cite{liu2019automatic} with PRSummarizer, who approached the challenge as a text summarization problem. They introduced a pointer generator to address out-of-vocabulary words \cite{see2017get}, and employed a reinforcement learning technique known as self-critical sequence training (SCST) \cite{rennie2017self} to construct a seq2seq model. Later, Fang et al. \cite{fang2022prhan} introduced a hybrid attention mechanism that enabled faster execution efficiency and better performance. 
Additionally, they employed a technique known as byte pair encoding \cite{gage1994new} to handle out-of-vocabulary words and effectively represent non-existing words in the vocabulary by combining sub-word units.%
Kuang et al. \cite{kuang2021automatic}, on the other hand,  tackled a different aspect of the generation problem, namely the issue of granularity. To generate improved descriptions for PRs with a larger number of commit messages, they modeled the task as an extractive text summarization problem rather than an abstractive one and employed a heterogeneous graph neural network to select words for the generated descriptions. Our work and the aforementioned works are complementary. These studies motivated us to construct a benchmark dataset and introduce new cleaning heuristics so that the description generation models can give better results.

\section{Conclusion} \label{conc-sec}
In this study, we propose four distinct heuristics to eliminate noise from PR description datasets. Using our heuristics, we construct a benchmark dataset consisting of 169K+ pull requests from 513 GitHub repositories. To assess our heuristics, we conducted automatic and manual evaluations on state-of-the-art summarization approaches for the PR description generation task. 
The experimental findings demonstrate that models trained on the dataset cleaned with our heuristics show average improvements of 8.6\%, 8.7\%, and 8.5\% in F1 scores for ROUGE-1, ROUGE-2, and ROUGE-L metrics, respectively, compared to those trained on the uncleaned dataset. Manual evaluation also shows that descriptions generated by BART, the best-performing model across the four, when trained on a cleaned dataset, are preferred over descriptions generated when trained on an unclean dataset. We believe that the findings of this work are essential for evaluating PR description generation models. These findings provide a solid foundation for improving model performance and the overall quality of automated PR descriptions.

\section{Data Availability}  \label{dataAvailability}
All datasets and code used in this study are openly available at \url{https://figshare.com/s/58ee9c2a4e9d951305d7}.

\bibliography{main}

\end{document}